\documentclass[prl,aps,twocolumn,showpacs]{revtex4}
\usepackage{amssymb}
\usepackage{amsmath}
\usepackage{graphicx}
\usepackage{wasysym}
\usepackage{color}

\usepackage{epstopdf}

\newcommand{\be}{\begin{eqnarray}}
\newcommand{\ee}{\end{eqnarray}}
\newcommand{\bbm}{\begin{bmatrix}}
\newcommand{\ebm}{\end{bmatrix}}
\renewcommand{\v}[1]{{\bf #1}}

\begin{document}
\title{Analytical Theory of Strongly Correlated Wigner Crystals in the Lowest Landau Level}

\author{Jun-Won \surname{Rhim}$^1$, Jainendra K. \surname{Jain}$^{2}$,  and Kwon \surname{Park}$^{1}$ }
\email[Electronic address:]{kpark@kias.re.kr}
\affiliation{$^1$School of Physics, Korea Institute for Advanced Study, Seoul 130-722, Korea}
\affiliation{$^2$Department of Physics, 104 Davey Lab, Pennsylvania State University, University Park, PA 16802}

\date{\today}

\begin{abstract}
In this work, we present an analytical theory of strongly correlated Wigner crystals (WCs) in the lowest Landau level (LLL) by constructing an approximate, but accurate effective two-body interaction for composite fermions (CFs) participating in the WCs. 
This requires integrating out the degrees of freedom of all surrounding CFs, which we accomplish analytically by approximating their wave functions by delta functions.
This method produces energies of various strongly correlated WCs that are in excellent agreement with those obtained from the Monte Carlo simulation of the full CF crystal wave functions.
We compute the compressibility of the strongly correlated WCs in the LLL and predict discontinuous changes at the phase boundaries separating different crystal phases. 
\end{abstract}



\maketitle

%%%%%%%%%%%%%%%%%%%%%%%%%%%%%%%%%%%%%%%%%%%%%%%%%%%%%%%%%%%%%%%%%%%%%%%%%

Wigner predicted more than eighty years ago~\cite{Wigner} that, when the interaction energy is dominant over the kinetic energy, electrons form a crystal, which is called WC after its originator. 
One possible way of suppressing the kinetic energy relative to the interaction energy is via the application of a strong magnetic field to two-dimensional electron systems~\cite{Lozovik}, which generates a fascinating series of various emergent quantum phases. 
The most celebrated examples are the fractional quantum Hall states~\cite{FQHE}, where new emergent quasiparticles called CFs form the quantum Hall liquid states~\cite{Jain89, Jain07}.
The quantum Hall liquid states are more effective in minimizing the interaction energy than WCs for a range of filling factors that is not too low. 
Nevertheless, WCs are expected to occur at sufficiently low filling factors.
Indeed, insulating states observed at filling factor $\nu <1/5$ are interpreted as pinned WCs~\cite{Andrei88, Jiang90, Goldman90, Williams91, Li91, Li97, Chen04, Engel97, Li00, Pan02, Ye02, Chen06, Sambandamurthy06, Shayegan_Review98, Shayegan_Review06}.
More recently, indications of the existence of a WC in the LLL have been seen through commensurability magneto-resistance  oscillations in bilayer Hall systems composed of a CF sea in one layer and a WC in the other~\cite{Liu14}.

Numerous theoretical studies have investigated the nature of WCs in the LLL~\cite{Maki83, Lam84, Levesque84, Esfarjani90, Cote91, Yi98, Narevich01, Yang01, Shibata03, Mandal03, Jeon04, Chang05, He05, Chang06, Shi07, Archer11}.
Initially, Maki and Zotos~\cite{Maki83} considered an uncorrelated Hartree-Fock WC of electrons, which was improved upon by Lam and Girvin~\cite{Lam84} by incorporating correlations.
In view of the success of the CF theory, Yi and Fertig~\cite{Yi98} proposed a strongly correlated WC 
composed of CFs, which was subsequently shown by Chang {\em et al.}~\cite{Chang05} to provide an accurate description at low filling factors ($\nu\leq1/5$).
Despite these extensive theoretical works, the calculation of a precise phase diagram of quantum Hall liquids versus CF crystals (CFCs) remained stalled for many years due to difficulties in obtaining the energy of CFCs in the thermodynamic limit accurately. 
This issue was resolved in a recent work~\cite{Archer13} inspired by the Thomson problem~\cite{Thomson04}. 
Here, the CFC wave functions are constructed in the spherical geometry by placing the WC wave packet centers at the locations that minimize the Coulomb energy of $N$ charged point particles on the surface of a sphere.  
Locally, these minimum energy positions resemble the hexagonal lattice, which is the minimum energy symmetry for a classical 2D electron crystal~\cite{Bonsall77}.
This allows a precise investigation of the CFC wave functions up to a fairly large system size ($N \sim 100$)~\cite{Archer13}.

The Monte Carlo (MC) simulation of the CFC wave functions is computationally quite expensive and rather difficult to implement.  Furthermore, it turns out that even though the energy obtained from this method enables a determination of the phase diagram, it is not sufficiently accurate to allow an evaluation of quantities such as compressibility, which is related to the second derivative of the energy. 
We develop in this work an analytical theory of the CFCs by constructing an accurate effective two-body interaction, which is based on the two-body wave function of CFs participating in the CFCs. This requires integrating out the degrees of freedom of all surrounding CFs, which we accomplish analytically by approximating their wave functions as delta functions.
We call this approach the ``renormalized two-body formalism," to be contrasted with the ``isolated two-body formalism" where the effects of all surrounding CFs are neglected.
The CFC energies obtained from the renormalized two-body formalism are in excellent agreement with those obtained from the MC simulation of the full CFC wave functions. 
With these analytical results, we obtain the compressibility and predict that its measurements, such as those carried out in GaAs heterostructures or in graphene~\cite{Eisenstein94, Ponomarenko10, Yu13, Martin10, Feldman12, Feldman13}, can detect the phase diagram of the CFCs.

We begin by constructing the wave function for the CFC carrying $2p$ vortices, or in short $^{2p}$CFC, as follows:
\begin{align}
\Psi^{^{2p}\mathrm{CFC}}_{\nu} = \prod_{j<k}(z_j-z_k)^{2p}\Psi^{\mathrm{MZ}}_{\nu^*} ,
\label{eq:CFC_wavefunc}
\end{align}
where $z_j=x_j +i y_j$ is the coordinates of the $j$-th electron. 
$\nu$ and $\nu^*$ denote the filling factors of electron and CFCs, respectively.  The function 
$\Psi^{\mathrm{MZ}}_{\nu^*} = \textrm{Det} \left[ \phi_{\v R_i}(\v r_j) \right]$ is the MZ wave function for the uncorrelated WC, comprising the LLL coherent-state wave function $\phi_{\v R}(\v r) = \frac{1}{\sqrt{2\pi}} \exp{\left[ -(\v r - \v R)^2/4 -i(xY - yX)/2 \right]}$ centered at $\v R=(X,Y)$ \cite{Maki83}.

Without the Jastrow factor correlation, the energy of the uncorrelated Maki-Zotos (MZ) WC, i.e., $^0$CFC, can be computed from the effective two-body interaction~\cite{Maki83}:
\begin{align}
\frac{V^{\mathrm{MZ}}(R_{ij})}{e^2/\epsilon l_{\rm B}} 
&= \left\langle \psi^{^{0}\mathrm{CFC}} \right| |\v r_1 - \v r_2|^{-1} \left| \psi^{^{0}\mathrm{CFC}} \right\rangle
\nonumber \\
&=\frac{\sqrt{\pi}}{4}\textrm{sech}{(R_{ij}^2/8)} I_0{(R_{ij}^2/8)} ,
\label{eq:MZ_int}
\end{align}
where $R_{ij} = |\v R_{ij}| = |\v R_i - \v R_j|$ is the distance between two crystal lattice centers.  
The uncorrelated two-body wave function is given by $\psi^{^{0}\mathrm{CFC}} (\v r_1, \v r_2) = {\cal C}_0 {\cal A} [ \phi_{\v R_i}(\v r_1)\phi_{\v R_j}(\v r_2) ]$ with ${\cal C}_0$ being the normalization constant and ${\cal A}$ being the antisymmetrization operator.
$I_0$ is the modified Bessel function of the first kind.
The energy per particle of the uncorrelated MZ WC, $E^{\rm MZ}$, 
is computed by performing the Madelung-type lattice summation of the MZ effective two-body interaction energy between all pairs of electrons in the hexagonal lattice:
\begin{align}
\frac{E^{\mathrm{MZ}}}{e^2/\epsilon l_{\rm B}} = \frac{1}{2N} \sum_{i \neq j} \left( \frac{V^{\mathrm{MZ}}(R_{ij})}{e^2/\epsilon l_{\rm B}} -\frac{1}{R_{ij}} \right)-\alpha \sqrt{\nu} ,
\label{eq:MZ_total_energy}
\end{align}
where the terms $\frac{1}{2N} \sum_{i \neq j} \frac{1}{R_{ij}} +\alpha\sqrt{\nu}$ with $\alpha=0.782133$ are subtracted to take into account the neutralizing effect of the uniform positive-charge background~\cite{Bonsall77}.

%%%%%%%%%%%%%%%%%%%%%%%%%%%%%%%%%%%%%%%%%%%%%%%%%%%%%%%%%%
\begin{figure}[t]
\includegraphics[width=0.9\columnwidth]{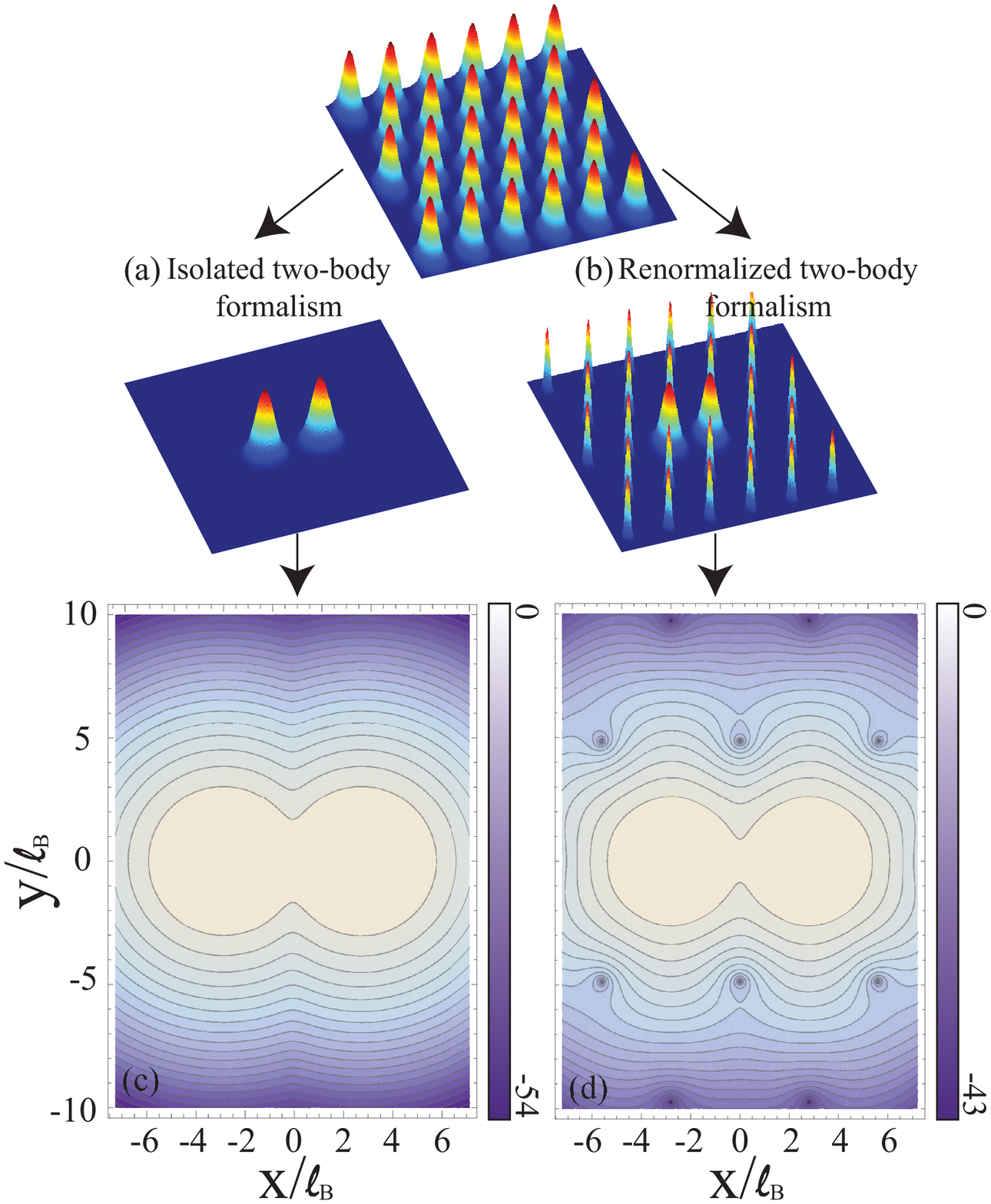}
\caption{
(Color online) 
Schematic diagram showing (a) the isolated and (b) the renormalized two-body formalism, accompanied by the probability density of the two-body CF wave function in each formalism, 
(c) $\rho_{\rm isol}(\v r) = \int d^2 \v r^\prime | \psi^{^{2p}{\rm CFC}}_{\rm isol}(\v r, \v r^\prime) |^2$
and 
(d) $\rho_{\rm renorm}(\v r) = \int d^2 \v r^\prime | \psi^{^{2p}{\rm CFC}}_{\rm renorm}(\v r, \v r^\prime) |^2$, 
respectively. 
Here, we set $2p=2$ and the nominal distance between crystal centers to be $6 l_{\rm B}$.
Note that probability densities are plotted in the natural log scale.}
\label{fig:Schematic}
\end{figure}
%%%%%%%%%%%%%%%%%%%%%%%%%%%%%%%%%%%%%%%%%%%%%%%%%%%%%%%%%%

For strongly correlated WCs, we need to take care of the Jastrow factor.
As a first try, we begin by focusing on two isolated CFs and ignoring all other surrounding CFs, in which situation the two-body CF wave function is obtained as
$\psi^{^{2p}\mathrm{CFC}}_{{\rm isol}}(\v r_1, \v r_2) = {\cal C}_{2p} (z_1 - z_2)^{2p} {\cal A}\left[ \phi_{\v R_i}(\v r_1)\phi_{\v R_j}(\v r_2) \right]$.
The normalization constant is ${\cal C}_{2p}= [ {_1F_1}(2p+1;1;R_{ij}^2/4) - {_1F_1}(2p+1;1;-R_{ij}^2/4) ]^{-1/2}/ [\pi 2^{2p+1} \sqrt{2\Gamma(2p+1)} e^{-R_{ij}^2/8} ]$ where ${_1F_1}(a;b;z)=\sum_{n=0}^\infty \frac{a^{(n)}}{b^{(n)}n!}z^n$ is the Kummer's hypergeometric function with $a^{(n)}=a(a+1)\cdots(a+n-1)$.
We refer to this approach the isolated two-body formalism. 
Figure~\ref{fig:Schematic}~(a) shows a schematic diagram for the isolated two-body formalism, which is accompanied by the probability density of the two-body CF wave function defined by $\rho_{\rm isol}(\v r) = \int d^2 \v r^\prime |\psi^{^{2p}\textrm{CFC}}_{\rm isol} (\v r, \v r^\prime)|^2$ in Fig.~\ref{fig:Schematic}~(c).

It is important to note that the actual distance between composite fermions $d_{ij}$ is not exactly equal to the nominal distance $R_{ij}$, because the Jastrow factor incorporates an additional repulsion into the two-body wave function, pushing the wave packets slightly farther apart.
To take this effect into account, we define $d_{ij}$ to be the \emph{median} distance between two maxima in the two-body probability density.
Specifically, $d_{ij}$ is related with $R_{ij}$ so that $\frac{\partial}{\partial \v r} |\psi^{^{2p}\mathrm{CFC}}_{\rm isol}(\v r_1 , \v r_2)|^2 =0$ at $\v r = \v r_1 - \v r_2 =(d_{ij},0)$ with $\v R_{ij}=(R_{ij}, 0)$, which, after some algebra, becomes $d_{ij}^2- R_{ij} d_{ij} \coth{(R_{ij} d_{ij} /4 )} -8p= 0$. 
Note that $d_{ij}$ is slightly larger than $R_{ij}$ with their difference growing as $2p$ increases. 
Given this information, the effective two-body interaction between CFs in the isolated two-body formalism is computed as follows:
\begin{align}
&\frac{V^{^{2p}\textrm{CFC}}_{\rm isol}(d_{ij})}{e^2/\epsilon l_{\rm B}} 
= \left\langle \psi^{^{2p}\textrm{CFC}}_{\rm isol} \right|  |\v r_1 -\v r_2|^{-1} \left| \psi^{^{2p}\textrm{CFC}}_{\rm isol} \right\rangle
\nonumber \\
&= {\cal B}_{2p} 
\frac{ {_1F_1}(2p+1/2;1;R_{ij}^2/4)  - {_1F_1}(2p+1/2;1;-R_{ij}^2/4) } 
{ L_{2p}(-R_{ij}^2/4)e^{R_{ij}^2/4} - L_{2p}(R_{ij}^2/4)e^{-R_{ij}^2/4} }
\label{eq:V_anti}
\end{align}
where ${\cal B}_{2p}=\Gamma(2p+1/2)/[2\Gamma(2p+1)]$ and $L_n(x)$ is the Laguerre polynomial.
For the hexagonal lattice, $d_{ij}$ is set equal to the distance between various crystal lattice centers via $d_{ij} = \sqrt{i^2 a^2+j^2 b^2}$ with $a=(4\pi/\sqrt{3}\nu)^{1/2}$ and $b=\sqrt{3}a$.

%%%%%%%%%%%%%%%%%%%%%%%%%%%%%%%%%%%%%%%%%%%%%%%%%%%%%%%%%%
\begin{figure}[t]
\includegraphics[width=0.8\columnwidth]{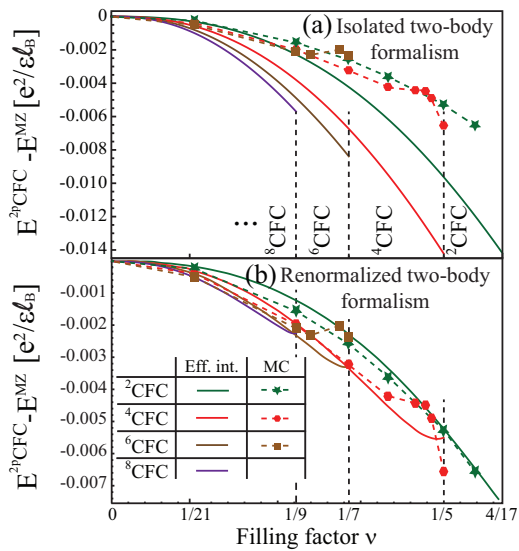}
\caption{
(Color online) Energy per particle of various CFC states, $E^{^{2p}{\rm CFC}}$, in reference to that of the MZ WC, $E^{\rm MZ}$, as a function of filling factor, $\nu$, obtained in (a) the isolated and (b) the renormalized two-body formalism. 
Symbols and dashed lines are obtained from the MC simulation of the full CFC wave functions.
Note that each $^{2p}$CFC is the lowest energy state (at least, among various CFC states) within the range of $1/(2p+3) < \nu \lesssim 1/(2p+1)$ for $2p \geq 2$.
The MZ WC, or $^0$CFC is not energetically favorable at any filling factor ranges. 
}
\label{fig:Energy}
\end{figure}
%%%%%%%%%%%%%%%%%%%%%%%%%%%%%%%%%%%%%%%%%%%%%%%%%%%%%%%%%%

In the isolated two-body formalism, the energy per particle of $^{2p}{\rm CFC}$, $E^{^{2p}\mathrm{CFC}}_{\rm isol}$, is evaluated similarly to Eq.~\eqref{eq:MZ_total_energy} by replacing $V^{\mathrm{MZ}}$ with $V^{^{2p}\textrm{CFC}}_{\rm isol}$.
Figure~\ref{fig:Energy} (a) shows $E^{^{2p}\mathrm{CFC}}_{\rm isol}-E^{\rm MZ}$ as a function of filling factor, 
which is compared with the MC simulation results obtained from the full CF wave function in the spherical geometry. 
As one can see, $E^{^{2p}\mathrm{CFC}}_{\rm isol}-E^{\rm MZ}$ shows a reasonably good agreement with the MC simulation results, especially at low filling factors.
There are, however, some sizable quantitative discrepancies in general.

To improve upon the isolated two-body formalism, it is necessary to include Jastrow-factor correlation effects arising from all surrounding CFs in some fashion.
To this end, we test an approximation called {\it the surrounding delta-function approximation}, where the wave functions of all surrounding CFs are approximated as delta functions. 
See Fig.~\ref{fig:Schematic}~(b) for a schematic diagram.
This approximation should be exact in the limit of large separation between crystalline CFs. 
Within the surrounding delta-function approximation, one can integrate out the degrees of freedom of all surrounding CFs and then derive the analytical two-body CF wave function.
For convenience, we call this approach the renormalized two-body formalism.

In the renormalized two-body formalism, the two-body CF wave function is written as 
$\psi^{^{2p}\textrm{CFC}}_\textrm{renorm} (\v r_1, \v r_2)
\propto  \psi^{^{2p}\textrm{CFC}}_\textrm{isol} (\v r_1, \v r_2) 
\prod_{k\neq i, j} (z_1- Z_k)^{2p}  \prod_{l\neq i, j} (z_2- Z_l)^{2p}$ ,
where $Z_k$ denotes the coordinates of the $k$-th crystal lattice center. 
After dividing a constant factor $\prod_{k\neq i,j}(Z_i-Z_k)^{2p} \prod_{l\neq i,j}(Z_j-Z_l)^{2p}$, The above equation can be rewritten as
\begin{align}
\psi^{^{2p}\textrm{CFC}}_\textrm{renorm} (\v r_1, \v r_2)
&= \tilde{{\cal C}}_{2p} (z_1-z_2)^{2p} 
\nonumber \\
&\times {\cal A} \left[  \Gamma_{ij}^{2p}(\v r_1, \v r_2)  \tilde{\phi}_{\v R_i}(\v r_1)  \tilde{\phi}_{\v R_j}(\v r_2) \right],
\label{eq:psi_renorm2}
\end{align}
where $\tilde{{\cal C}}_{2p}$ is the normalization constant,
$\Gamma_{ij}^{2p}(\v r_1, \v r_2) = (Z_i-Z_j)^{4p}/[(z_1-Z_j)^{2p} (z_2-Z_i)^{2p}]$, and
$\tilde{\phi}_{\v R_i}(\v r) 
= \phi_{\v R_i}(\v r) \prod_{k\neq i} \frac{(z-Z_k)^{2p}}{(Z_i-Z_k)^{2p}} \equiv \phi_{\v R_i}(\v r) \left[ {\cal F}_{\v R_i}(\v r) \right]^p$, 
which is the renormalized version of the coherent-state wave function centered at $\v R_i$.

It is important to note that all the complicated many-body correlations are embedded in the renormalization factor ${\cal F}_{\v R_i}(\v r)$.
The success of this approach stems from the fact that we are able to obtain the analytical form of ${\cal F}_{\v R_i}(\v r)$:
\begin{align}
{\cal F}_{\v R_i}(\v r) 
&=\frac{ \theta_1 \left( \frac{\pi}{a}(z-Z_i) | i \frac{b}{a} \right)  \theta_1\left( \frac{i\pi}{b}(z-Z_i) | i \frac{a}{b} \right) } 
{i \frac{\pi^2}{ab} \theta^\prime_1\left(0 | i \frac{b}{a} \right) \theta^\prime_1\left(0 | i \frac{a}{b} \right)  (z-Z_i)^2}
\nonumber \\
&\times \frac{ \theta_3\left( \frac{\pi}{a}(z-Z_i) | i \frac{b}{a} \right)  \theta_3\left( \frac{i\pi}{b}(z-Z_i) | i \frac{a}{b} \right) } 
{\theta_3\left(0 | i \frac{b}{a} \right) \theta_3\left(0 | i \frac{a}{b} \right)} ,
\label{eq:renorm_factor}
\end{align}
where $\theta_n(z|\tau) $ is the Jacobi theta function and $b=\sqrt{3}a$ with $a$ being the lattice constant.
See Supplementary Material for details. 
Figure~\ref{fig:Schematic}~(d) shows the renormalized probability density defined by $\rho_{\rm renorm}(\v r) = \int d^2 \v r^\prime |\psi^{^{2p}\textrm{CFC}}_{\rm renorm} (\v r, \v r^\prime)|^2$.
As one can see, the renormalized probability density exhibits the hexagonal symmetry of the WC which was absent in the isolated two-body formalism.

As before, the energy per particle in the renormalized two-body formalism, $E^{^{2p}\mathrm{CFC}}_{\rm renorm}$, can be computed as the Madelung-type lattice summation of the renormalized effective two-body interaction, 
$V^{^{2p}\textrm{CFC}}_{\rm renorm}/(e^2/\epsilon l_{\rm B}) = \langle \psi^{^{2p}\textrm{CFC}}_{\rm renorm}| |\v r_1 -\v r_2|^{-1} | \psi^{^{2p}\textrm{CFC}}_{\rm renorm} \rangle$. 
Note that, in general, $V^{^{2p}\textrm{CFC}}_{\rm renorm}$ depends on the vector $\v R_{ij}$, not just on the distance $R_{ij}$.
Conveniently, however, $V^{^{2p}\textrm{CFC}}_{\rm renorm}$ can be well approximated as a function of only $R_{ij}$ if $R_{ij} \gtrsim [4\pi(2p+1)/\sqrt{3}]^{1/2}$.
Also, due to the additional repulsion from all surrounding CFs, the actual distance $d_{ij}$ becomes quite close to the nominal distance $R_{ij}$ in most situations so that $R_{ij}$ can be simply regarded as $d_{ij}$.

Figure~\ref{fig:Energy}~(b) shows $E^{^{2p}\mathrm{CFC}}_{\rm renorm} - E^{\rm MZ}$ as a function of filling factor.
As one can see, the results from the renormalized two-body formalism are in excellent agreement with those from the MC simulation of the full CFC wave functions. 
It is interesting to observe that the renormalized two-body formalism can even capture the initial upturn of the energy near $\nu=1/(2p+1)$ for each corresponding $^{2p}$CFC.
The most significant discrepancy is that the MC results exhibit sharp drops immediately following such upturns. 
It is important to note, however, that, regardless of being isolated or renormalized, the two-body formalism for each $^{2p}$CFC is supposed to lose its validity near $\nu=1/(2p+1)$ since, here, wave packets are highly overlapping and thus higher-body corrections become important. 
Given the simplification in the two-body formalism, we consider the agreement to be excellent.

%%%%%%%%%%%%%%%%%%%%%%%%%%%%%%%%%%%%%%%%%%%%%%%%%%%%%%%%%%
\begin{figure}[t]
\includegraphics[width=0.8\columnwidth]{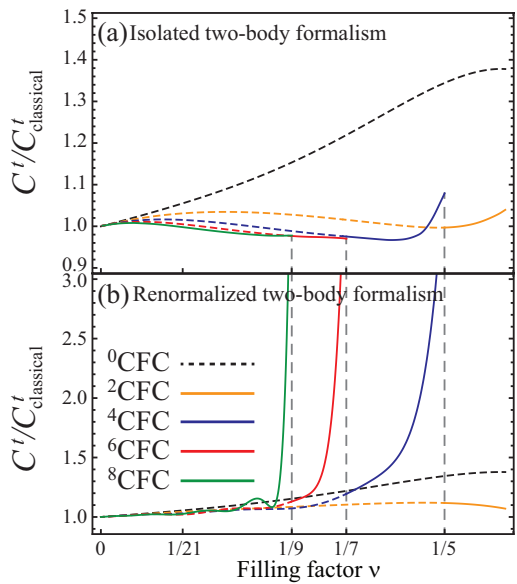}
\caption{
(Color online) Shear modulus, $C^t$, of various CFC states as a function of filling factor, $\nu$, obtained in (a) the isolated and (b) the renormalized two-body formalism. 
For convenience, $C^t$ is expressed as its relative ratio with respect to that for the classical WC, $C^t_{\mathrm{classical}}/(e^2/\epsilon l_B)=0.0978 \sqrt{\nu}$.
Note that $C^t$ of each $^{2p}$CFC is valid only within the range of $1/(2p+3) < \nu \lesssim 1/(2p+1)$, where its curve is plotted in a solid line, denoting that the $^{2p}$CFC is the lowest energy state here.
}
\label{fig:Shear}
\end{figure}
%%%%%%%%%%%%%%%%%%%%%%%%%%%%%%%%%%%%%%%%%%%%%%%%%%%%%%%%%%

Bolstered by the quantitative accuracy of the renormalized two-body formalism, we now compute the shear modulus of CFC states, $C^t$, as a function of filling factor. 
To this end, we utilize the following relation~\cite{Archer13}:
$C^t = \frac{1}{2}\nu^2\frac{\partial^2}{\partial \nu^2} E^{^{2p}\mathrm{CFC}}.$ 
It is important to note that this relation is derived under the assumption that only the two-body interaction is relevant, and is therefore consistent with our two-body formalism.
Figure~\ref{fig:Shear} shows $C^t$ of various CFC states as a function of filling factor obtained by using $E^{^{2p}\mathrm{CFC}}_{\rm isol}$ and $E^{^{2p}\mathrm{CFC}}_{\rm renorm}$. 
While $C^t$ obtained from the isolated two-body formalism is not to be trusted, it shows the relative importance of correlations from all surrounding CFs, which enhance $C^t$ significantly.
In particular, $C^t$ in the renormalized two-body formalism shows a series of huge enhancements followed by discontinuous drops near $\nu=1/(2p+1)$, which can be used as a distinctive signature for a phase transition between different CFC states.

%%%%%%%%%%%%%%%%%%%%%%%%%%%%%%%%%%%%%%%%%%%%%%%%%%%%%%%%%%
\begin{figure}[t]
\includegraphics[width=0.8\columnwidth]{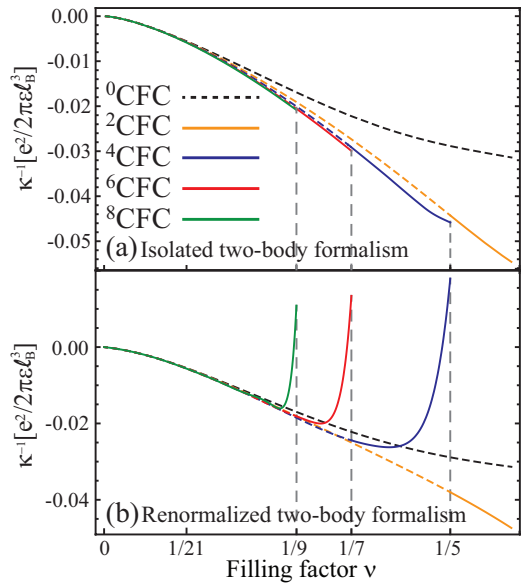}
\caption{
(Color online) Inverse of the compressibility, $\kappa^{-1}$, of various CFC states as a function of filling factor, $\nu$, obtained in (a) the isolated and (b) the renormalized two-body formalism.
Similar to $C^t$, $\kappa^{-1}$ of each $^{2p}$CFC is valid only within the range of $1/(2p+3) < \nu \lesssim 1/(2p+1)$, where its curve is plotted in a solid line.
}
\label{fig:InverseComp}
\end{figure}
%%%%%%%%%%%%%%%%%%%%%%%%%%%%%%%%%%%%%%%%%%%%%%%%%%%%%%%%%%

Another important observable is the compressibility, whose inverse can be computed as follows~\cite{Mahan}:
$\kappa^{-1} = \frac{1}{2\pi l^2_B} \nu^2 \frac{\partial^2}{\partial \nu^2} (\nu E^{^{2p}{\rm CFC}}),$ 
where it is used that the electron density is related with the filling factor via $n=\nu/2\pi l^2_B$.
Figure~\ref{fig:InverseComp} shows $\kappa^{-1}$ of various CFC states as a function of filling factor. 
At first sight, it may seem strange that the compressibility becomes negative in some regimes. 
This does not, however, mean an instability here since, by construction, we do not allow the positive background charge to relax.  What we obtain above is the electronic part of the compressibility called the {\it proper} compressibility~\cite{Giuliani}, which can be negative.  In fact, the proper compressibility is directly measured in capacitive experiments~\cite{Eisenstein94, Ponomarenko10, Yu13} or by scanning single-electron transistor~\cite{Martin10, Feldman12, Feldman13}. 
Compressibility has served as a powerful tool for detecting phase transitions between different fractional quantum Hall states as well as between differently spin polarized states at a given fraction~\cite{Martin10, Feldman12, Feldman13}. 
Our calculations predict discontinuous changes in compressibility at the phase boundaries separating the different CFC phases, which can allow a determination of the phase diagram.  
Observation of such transitions inside the crystal phase will serve as direct evidence for the correlated CF character of WCs in the LLL, corroborating existing experimental indications of the CFC states~\cite{Hatke14, Liu14_PRL}.

It is noteworthy that this is the first example where an accurate analytical treatment has become possible for a strongly correlated state in the LLL. 
It would be worth investigating if our method can be extended to the liquid states of CFs.

%%%%%%%%%%%
% Acknowledgment %
%%%%%%%%%%%

The authors are grateful to Alexander C. Archer for sharing the MC simulation results for the energy of CFC states (from Ref.~\cite{Archer13}).
JKJ acknowledges financial support from the US National Science Foundation.

%%%%%%%%%%%%%%%
% Supplementary Material %
%%%%%%%%%%%%%%%

\begin{widetext}

\vspace{5mm}
\begin{center}
{\bf Supplementary Material: Theta-function Representation of the Renormalization Factor}
\end{center}
\vspace{5mm}

The goal of this Supplementary Material is to evaluate the renormalization factor ${\cal F}_{\v R_i}(\v r)$, which is defined as follows:
\begin{align}
{\cal F}_{\v R_i}(\v r) = \prod_{k\neq i} \frac{(z-Z_k)^{2}}{(Z_i-Z_k)^{2}}  ,
\label{eq:renorm_factor}
\end{align}
where the product index, $k$, goes over all crystal lattice points on the triangular lattice except the $i$-th. 
Now, for a reason that becomes clear below, it is convenient to decompose the triangular lattice into two overlapping rectangular lattices, say, the rectangular lattice 1 (RL$_1$) and 2 (RL$_2$). 
See Fig.~\ref{fig:lattice_decom} for illustration.

%%%%%%%%%%%%%%%%%%%%%%%%%%%%%%%%%%%%%%%%%%%%%%%%%%%%%%%%%%
\begin{figure}[b]
\includegraphics[width=0.6\columnwidth]{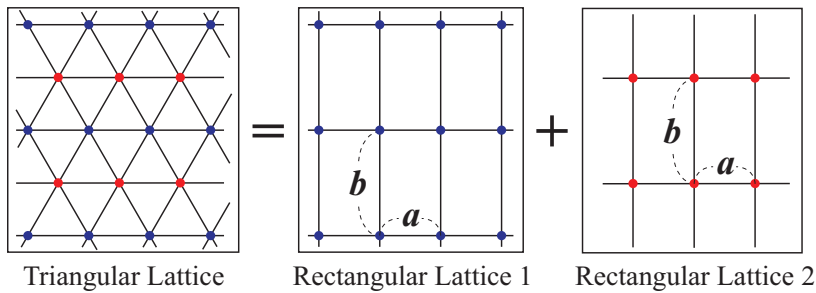}
\caption{
(Color online) 
Decomposition of the triangular lattice into two overlapping rectangular lattices, say, the rectangular lattice 1 (RL$_1$) and 2 (RL$_2$).
Note that $b=\sqrt{3}a$.
}
\label{fig:lattice_decom}
\end{figure}
%%%%%%%%%%%%%%%%%%%%%%%%%%%%%%%%%%%%%%%%%%%%%%%%%%%%%%%%%%

By using this decomposition, Eq.~\eqref{eq:renorm_factor} can be rewritten as follows:
\begin{align}
{\cal F}_{\v R_i}(\v r) = {\cal F}^{\textrm{RL}_1}_{\v R_i}(\v r) {\cal F}^{\textrm{RL}_2}_{\v R_i}(\v r) ,
\label{eq:renorm_factor_decom}
\end{align}
where ${\cal F}^{\textrm{RL}_1}_{\v R_i}(\v r)$ and ${\cal F}^{\textrm{RL}_2}_{\v R_i}(\v r)$ are the renormalization factors containing contributions from RL$_1$ and RL$_2$, respectively:
\begin{align}
{\cal F}^{\textrm{RL}_\alpha}_{\v R_i}(\v r) = \prod_{k\neq i, \in \textrm{RL}_\alpha} \frac{(z-Z_k)^{2}}{(Z_i-Z_k)^{2}}  ,
\label{eq:renorm_factor_decom2}
\end{align}
where $\alpha=1$ or $2$.
Specifically, the coordinates of crystal lattice points can be specified via $Z^{\textrm{RL}_1}_{(m,n)}= ma +inb$ for RL$_1$ and $Z^{\textrm{RL}_2}_{(m,n)}=(m-1/2)a +i(n-1/2)b$ for RL$_2$, where $m$ and $n$ are integers, and $a$ and $b$ $(=\sqrt{3}a)$ are the lattice constants of the rectangular lattices along the $x$ and $y$ direction, respectively. 
In this notation, Eq.~\eqref{eq:renorm_factor_decom2} can be written as follows:
\begin{align}
{\cal F}^{\textrm{RL}_\alpha}_{\v R_i}(\v r) &= \prod_{m, n=-\infty \atop (m,n) \neq {\bf R}_i}^{\infty} \left( 1 - \frac{z-Z_i}{Z^{\textrm{RL}_\alpha}_{(m,n)}-Z_i} \right)^2 ,
\label{eq:renorm_factor_decom3}
\end{align}
which can be further simplified as follows by redefining $(m,n)$ such that the coordinates of all crystal lattice points in the product are measured with reference to ${\bf R}_i$:
\begin{align}
{\cal F}^{\textrm{RL}_1}_{\v R_i}(\v r) &= \prod_{m, n=-\infty \atop (m,n) \neq (0,0) }^{\infty} \left( 1 - \frac{z-Z_i}{ma+inb} \right)^2 ,
\label{eq:renorm_factor_decom5}
\\
{\cal F}^{\textrm{RL}_2}_{\v R_i}(\v r) &= \prod_{m, n=-\infty}^{\infty} \left( 1 - \frac{z-Z_i}{(m-1/2)a+i(n-1/2)b} \right)^2 .
\label{eq:renorm_factor_decom6}
\end{align}

The concrete evaluation of the renormalization factor begins with the following, curious representation of the Jacobi theta function in terms of the double infinite product~\cite{NIST_Handbook}:
\begin{align}
\theta_1(z|\tau) &= z \theta^\prime_1(0|\tau) 
\lim_{N\rightarrow \infty} \prod_{n=-N}^N 
\lim_{M\rightarrow \infty} \prod_{m=-M \atop (n,m)\neq (0,0)}^M \left( 1+ \frac{z}{(m+n\tau)\pi }\right) ,
\label{eq:theta1}
\\
\theta_3(z|\tau) &= \theta_3(0|\tau) 
\lim_{N\rightarrow \infty} \prod_{n=1-N}^N 
\lim_{M\rightarrow \infty} \prod_{m=1-M}^M 
\left( 1+ \frac{z}{\left[m-\frac{1}{2}+\left(n-\frac{1}{2}\right)\tau\right]\pi }\right) ,
\label{eq:theta3}
\end{align}
where the order of the limits cannot be switched since the double products above are not absolutely convergent.   
Note that the Jacobi theta functions, $\theta_1(z|\tau)$ and $\theta_3(z|\tau)$, are defined as follows:
\begin{align}
\theta_1(z|\tau) &= \theta_1(z,q) = 2 \sum_{n=0}^{\infty} (-1)^n q^{(n+1/2)^2} \sin{((2n+1)z)} ,
\label{eq:theta1_def}
\\
\theta_3(z|\tau) &= \theta_3(z,q) =  1+ 2\sum_{n=1}^{\infty} q^{n^2} \cos{(2nz)} ,
\label{eq:theta3_def}
\end{align}
where $q=e^{i\pi\tau}$.

As one can see, Eqs.~\eqref{eq:renorm_factor_decom5} and \eqref{eq:renorm_factor_decom6} are very similar to Eqs.~\eqref{eq:theta1} and \eqref{eq:theta3}, respectively, which suggests the following, possible equalities:
\begin{align}
{\cal F}^{\textrm{RL}_1}_{x-{\rm prior}, \v R_i}(\v r) 
\stackrel{?}{=} \lim_{N\rightarrow \infty} \prod_{n=-N}^N 
\lim_{M\rightarrow \infty} \prod_{m=-M \atop (n,m)\neq (0,0)}^M 
\left( 1+ \frac{\frac{\pi}{a}(z-Z_i)}{\left(m+n i\frac{b}{a}\right)\pi }\right)^2 
= \left[ \frac{ \theta_1\left( \frac{\pi}{a}(z-Z_i) | i \frac{b}{a} \right)}{ \theta^\prime_1\left(0 | i\frac{b}{a}\right)  \frac{\pi}{a}(z-Z_i)} \right]^2,
\label{eq:renorm_factor_decom7}
\end{align}
and
\begin{align}
{\cal F}^{\textrm{RL}_2}_{x-{\rm prior}, \v R_i}(\v r) 
\stackrel{?}{=} 
\lim_{N\rightarrow \infty} \prod_{n=-N}^N 
\lim_{M\rightarrow \infty} \prod_{m=-M}^M 
\left( 1+ \frac{\frac{\pi}{a}(z-Z_i)}{\left[m-\frac{1}{2}+(n-\frac{1}{2}) i\frac{b}{a}\right]\pi }\right)^2 
= \left[ \frac{ \theta_3\left( \frac{\pi}{a}(z-Z_i) | i \frac{b}{a} \right) }{ \theta_3\left(0 |i\frac{b}{a}\right) } \right]^2,
\label{eq:renorm_factor_decom8}
\end{align}
where dummy product indices, $m$ and $n$, are reversed in sign and shifted by constant integers when necessary.
Despite the obvious similarity, however, there is a serious problem, which makes the above equalities ill-defined, or at least incorrect in the context that we intend. 
The problem stems from the previously-mentioned fact that the double products in the representation of the Jacobi theta function are not absolutely convergent and thus the limits should be taken with care.   
Technically, Eqs.~\eqref{eq:renorm_factor_decom7} and \eqref{eq:renorm_factor_decom8} can be regarded as being obtained by first multiplying the entire Jastrow factors contributed by the composite fermions located along the $x$ direction at a given $y$ coordinate and then repeating the same procedure for different $y$ coordinates to cover the whole two-dimensional lattice. 
For convenience, let us call this the {\it $x$-priority limit} procedure.

Another way of covering the whole two-dimensional lattice is to multiply the Jastrow factors along the $y$ direction at a given $x$ coordinate first and then repeating the same procedure for different $x$ coordinates. 
This alternative covering can be implemented via a slight rearrangement of the product, generating the following, possible equalities: 
\begin{align}
{\cal F}^{\textrm{RL}_1}_{y-{\rm prior}, \v R_i}(\v r) 
\stackrel{?}{=} \lim_{N\rightarrow \infty} \prod_{n=-N}^N 
\lim_{M\rightarrow \infty} \prod_{m=-M \atop (n,m)\neq (0,0)}^M 
\left( 1+ \frac{\frac{i\pi}{b}(z-Z_i)}{\left(n+m i\frac{a}{b}\right)\pi }\right)^2 
= \left[ \frac{ \theta_1\left( \frac{i\pi}{b}(z-Z_i) | i \frac{a}{b} \right) }{ \theta^\prime_1\left(0 |i\frac{a}{b}\right)  \frac{i\pi}{b}(z-Z_i) } \right]^2 ,
\label{eq:renorm_factor_decom9}
\end{align}
and
\begin{align}
{\cal F}^{\textrm{RL}_2}_{y-{\rm prior}, \v R_i}(\v r) 
\stackrel{?}{=} 
\lim_{N\rightarrow \infty} \prod_{n=-N}^N 
\lim_{M\rightarrow \infty} \prod_{m=-M}^M 
\left( 1+ \frac{\frac{i\pi}{b}(z-Z_i)}{\left[n-\frac{1}{2}+(m-\frac{1}{2}) i\frac{a}{b}\right]\pi }\right)^2 
= \left[ \frac{ \theta_3\left( \frac{i\pi}{b}(z-Z_i) | i \frac{a}{b} \right) }{ \theta_3\left(0 |i\frac{a}{b}\right) } \right]^2,
\label{eq:renorm_factor_decom10}
\end{align}
where, as before, dummy product indices, $m$ and $n$, are reversed in sign and shifted by constant integers when necessary.
In contrast to the $x$-priority limit procedure, we call this the {\it $y$-priority limit} procedure. 
It is intriguing to check numerically that Eqs.~\eqref{eq:renorm_factor_decom9} and \eqref{eq:renorm_factor_decom10} are actually different from Eqs.~\eqref{eq:renorm_factor_decom7} and \eqref{eq:renorm_factor_decom8}, respectively, which presents us a dilemma regarding what form to choose.

In fact, the most natural physical limit is neither the $x$ nor $y$-priority limit, but rather to evaluate the Jastrow-factor product for a given finite two-dimensional (2D) cluster with the linear size $S$ and increase $S$ while keeping the aspect ratio of the cluster.  
That is to say,
\begin{align}
{\cal F}^{\textrm{RL}_1}_{{\rm 2D}, \v R_i}(\v r) 
&\stackrel{!}{=} \lim_{S\rightarrow \infty} 
\prod_{n=-N_S}^{N_S} \prod_{m=-M_S \atop (n,m)\neq (0,0)}^{M_S} 
\left( 1- \frac{z-Z_i}{ma+inb}\right)^2 ,
\label{eq:renorm_factor1_2D} 
\\
{\cal F}^{\textrm{RL}_2}_{{\rm 2D}, \v R_i}(\v r) 
&\stackrel{!}{=} \lim_{S\rightarrow \infty} 
\prod_{n=-N_S}^{N_S} \prod_{m=-M_S}^{M_S} 
\left( 1- \frac{z-Z_i}{\left(m-\frac{1}{2}\right)a+i\left(n-\frac{1}{2}\right)b }\right)^2  ,
\label{eq:renorm_factor2_2D}
\end{align}
where $M_S=S$ and $N_S=\textrm{round}(\frac{a}{b}S)$.
Let us call this the {\it proper 2D limit} procedure.

%%%%%%%%%%%%%%%%%%%%%%%%%%%%%%%%%%%%%%%%%%%%%%%%%%%%%%%%%%
\begin{figure}[t]
\includegraphics[width=0.8\columnwidth]{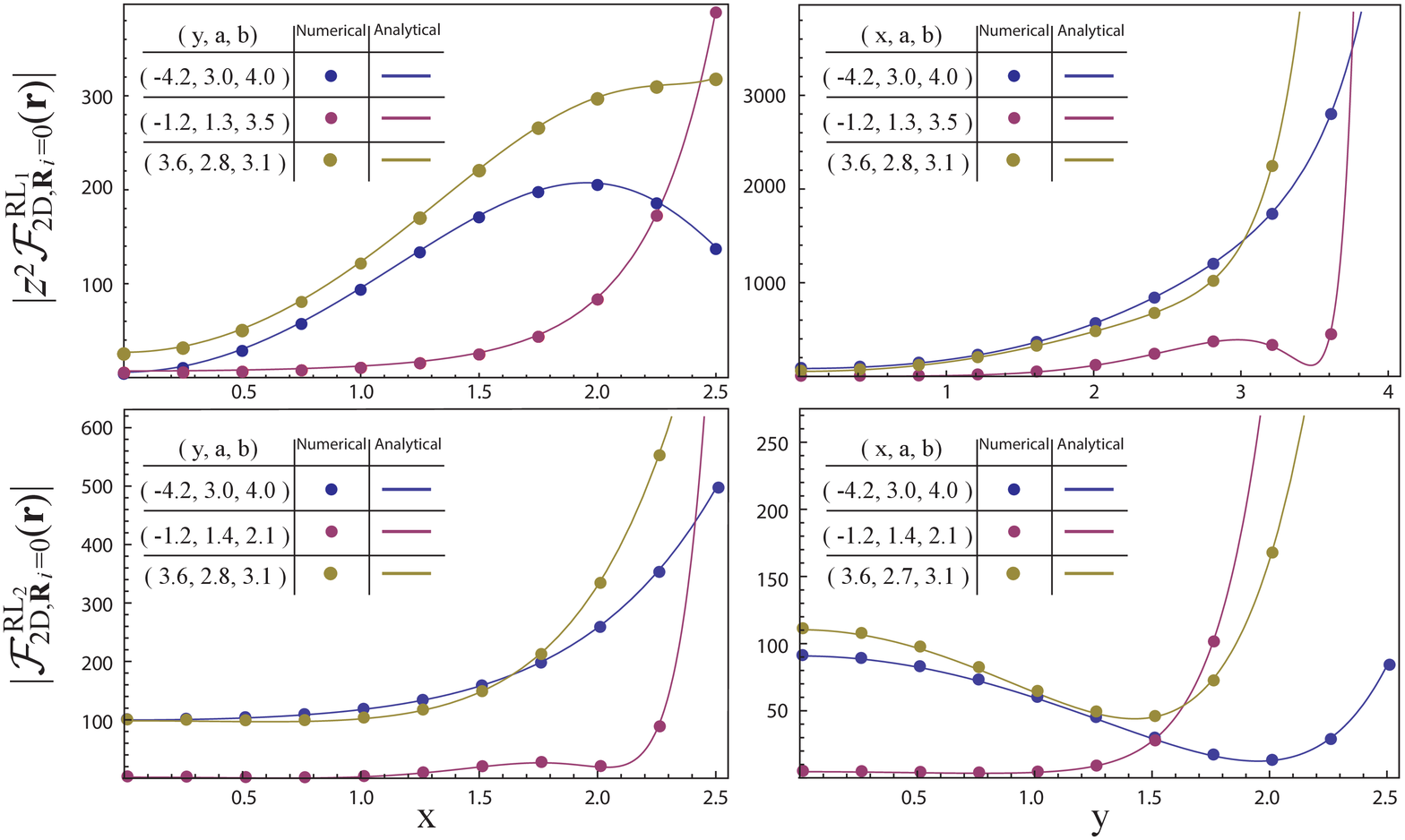}
\caption{
(Color online) Numerical tests for the validity of the geometric mean conjecture for the renormalization factor, ${\cal F}^{\textrm{RL}_\alpha}_{{\rm 2D}, \v R_i}(\v r)$. 
Here, $\v R_i$ is set to be at the origin, i.~e., $\v R_i=0$, without loss of generality.
We plot $|z^2 {\cal F}^{\textrm{RL}_1}_{{\rm 2D}, \v R_i=0}(\v r)|$ in the upper panels and $|{\cal F}^{\textrm{RL}_2}_{{\rm 2D}, \v R_i=0}(\v r)|$ in the bottom panels.
As one can see, the geometric mean conjecture for the renormalization factor yields essentially exact results in a wide range of ${\v r}=(x,y)$ for various values of the lattice constants, $a$ and $b$.
}
\label{fig:theta_test}
\end{figure}
%%%%%%%%%%%%%%%%%%%%%%%%%%%%%%%%%%%%%%%%%%%%%%%%%%%%%%%%%%

Now, a big question is how to actually evaluate the renormalization factor in this proper 2D limit.
To this end, we make a simple conjecture, which we call the {\it geometric mean conjecture}. 
Specifically, 
considering that the $x$-priority limit [in Eqs.~\eqref{eq:renorm_factor_decom7} and \eqref{eq:renorm_factor_decom8}] and the $y$-priority limit [in \eqref{eq:renorm_factor_decom9} and \eqref{eq:renorm_factor_decom10}] are the two opposite limits with reversed roles between the $x$ and $y$ direction, we make a conjecture that the proper 2D limit may be obtained by simply taking the geometric mean between the $x$ and $y$-priority limits: 
\begin{align}
{\cal F}^{\textrm{RL}_\alpha}_{\textrm{2D}, \v R_i} (\v r) = \sqrt{ {\cal F}^{\textrm{RL}_\alpha}_{x-\textrm{prior}, \v R_i}(\v r)  {\cal F}^{\textrm{RL}_\alpha}_{y-\textrm{prior}, \v R_i} (\v r)  } ,
\end{align}
which results in
\begin{align}
{\cal F}^{\textrm{RL}_1}_{\textrm{2D}, {\v R_i}}(\v r) = \frac{ 
\theta_1\left( \frac{\pi}{a}(z-Z_i) | i \frac{b}{a} \right)  \theta_1\left( \frac{i\pi}{b}(z-Z_i) | i \frac{a}{b} \right) 
} 
{i \frac{\pi^2}{ab} \theta^\prime_1\left(0 | i\frac{b}{a}\right) \theta^\prime_1\left(0 |i\frac{a}{b}\right)  (z-Z_i)^2
} ,
\label{eq:geomean_conjecture1}
\end{align}
and
\begin{align}
{\cal F}^{\textrm{RL}_2}_{\textrm{2D}, {\v R_i}}(\v r) = \frac{ 
\theta_3\left( \frac{\pi}{a}(z-Z_i) | i \frac{b}{a} \right)  \theta_3\left( \frac{i\pi}{b}(z-Z_i) | i \frac{a}{b} \right) 
} 
{\theta_3\left(0 | i\frac{b}{a}\right) \theta_3\left(0 |i\frac{a}{b}\right)
} ,
\label{eq:geomean_conjecture2}
\end{align}
where we take the liberty of ignoring the phase-factor issue, which may occur when taking the square root.  
We test the validity of the geometric mean conjecture by comparing Eqs.~\eqref{eq:geomean_conjecture1} and \eqref{eq:geomean_conjecture2} with the numerical results evaluated according to Eqs.~\eqref{eq:renorm_factor1_2D} and \eqref{eq:renorm_factor2_2D}, respectively, for sufficiently large $S$. 
Figure~\ref{fig:theta_test} shows the comparison between the geometric mean conjecture and the numerical results in a wide range of $\v r$ for various values of the lattice constants, $a$ and $b$.
As one can see, the agreement is essentially exact within numerical accuracy, establishing the validity of the geometric mean conjecture.

In conclusion, we obtain the final form of the renormalization factor as follows:
\begin{align}
{\cal F}_{\v R_i}(\v r) &= {\cal F}^{\textrm{RL}_1}_{{\rm 2D}, \v R_i}(\v r) {\cal F}^{\textrm{RL}_2}_{{\rm 2D}, \v R_i}(\v r)
= \frac{ 
\theta_1 \left( \frac{\pi}{a}(z-Z_i) | i \frac{b}{a} \right)  \theta_1 \left( \frac{i\pi}{b}(z-Z_i) | i \frac{a}{b} \right) 
} 
{i \frac{\pi^2}{ab} \theta^\prime_1\left(0 | i\frac{b}{a}\right) \theta^\prime_1\left(0 |i\frac{a}{b}\right) (z-Z_i)^2
}
%\cdot
\frac{ 
\theta_3\left( \frac{\pi}{a}(z-Z_i) | i \frac{b}{a} \right)  \theta_3\left( \frac{i\pi}{b}(z-Z_i) | i \frac{a}{b} \right) 
} 
{
\theta_3\left(0 | i\frac{b}{a}\right) \theta_3\left(0 |i\frac{a}{b}\right) 
}.
\end{align}

\end{widetext}

%%%%%%%%%%%%%%%%%%%%%%%%%%%%%%%%%%%%%%%%%%%%%%%%%%%%%%%%%%%%%%%%%%%%%%%%%


\begin{references}
%%%%%%%%%%%%%%%%%%%%%%%%%%%%%%%%%%%%%%%%%%%%%%%%%%%%%%%%%%%%%%%%%%%%%%%%%

\bibitem{Wigner} E. Wigner, Phys. Rev. {\bf 46}, 1002 (1934).

\bibitem{Lozovik} Y. E. Lozovik and V. I. Yudson, JETP Lett. {\bf 22}, 11 (1975).

\bibitem{FQHE} D. C. Tsui, H. L. Stormer, and A. C. Gossard, Phys. Rev. Lett. {\bf 48}, 1559 (1982).

\bibitem{Jain89} J. K. Jain, Phys. Rev. Lett. {\bf 63}, 199 (1989).

\bibitem{Jain07} For a review, see J. K. Jain, {\it Composite Fermions} (Cambridge University Press, Cambridge, 2007).



%%%%%%%%%%%%% Experimental evidence of pinned WC in the LLL %%%%%%%%%%%%%%%

%% Detection of a gapless magnetophonon excitation branch with radio frequency spectroscopy %%
\bibitem{Andrei88} E. Y. Andrei, G. Deville, D. C. Glattli, F. I. B. Williams, E. Paris, and B. Etienne, Phys. Rev. Lett. {\bf 60}, 2765 (1988).

%% Reentrant insulating behavior below and above nu=1/5 %%
\bibitem{Jiang90} H. W. Jiang, R. L. Willett, H. L. Stormer, D. C. Tsui, L. N. Pfeiffer, and K. W. West, Phys. Rev. Lett. {\bf 65}, 633 (1990);
H. W. Jiang, H. L. Stormer, D. C. Tsui, L. N. Pfeiffer, and K. W. West, Phys. Rev. B {\bf 44}, 8107 (1991).

%% Electric-field threshold conduction %%
\bibitem{Goldman90} V. J. Goldman, M. Santos, M. Shayegan, and J. E. Cunningham, Phys. Rev. Lett. {\bf 65}, 2189 (1990).

%% Conduction threshold and pinning frequency of magnetically induced Wigner solid %%
\bibitem{Williams91} F. I. B. Williams, P. A. Wright, R. G. Clark, E. Y. Andrei, G. Deville, D. C. Glattli, O. Probst, B. Etienne, C. Dorin, C. T. Foxon, and J. J. Harris, Phys. Rev. Lett. {\bf 66}, 3286 (1991).

%% Low-frequency noise in the reentrant insulating phase around the 1/5 FQH liquid %%
\bibitem{Li91} Y. P. Li, T. Sajoto, L. W. Engel, D. C. Tsui, and M. Shayegan, Phys. Rev. Lett. {\bf 67}, 1630 (1991).

%% Microwave conductivity resonance of 2D hole system %%
\bibitem{Li97} C.-C. Li, L. W. Engel, D. Shahar, D. C. Tsui, and M. Shayegan, Phys. Rev. Lett. {\bf 79}, 1353 (1997).

%% Microwave resonance in low-filling phases of two-dimensional electron and hole systems %%
\bibitem{Engel97} L. W. Engel, C.-C. Li, D. Shahar, D. C. Tsui, and M. Shayegan, Physica E {\bf 1}, 111 (1997).

%% Microwave resonance and weak pinning in two-dimensional hole systems at high magnetic fields %%
\bibitem{Li00} C. C. Li, J. Yoon, L. W. Engel, D. Shahar, D. C. Tsui, and M. Shayegan, Phys. Rev. B {\bf 61}, 10905 (2000).

%% Transition from an electron solid to the sequence of fractional quantum Hall states at very low Landau level filling factor: melting of WC to FQHS %%
\bibitem{Pan02} W. Pan, H. L. Stormer, D. C. Tsui, L. N. Pfeiffer, K. W. Baldwin, and K. W. West, Phys. Rev. Lett. {\bf 88}, 176802 (2002).

%% Correlation lengths of the Wigner-crystal order in a two-dimensional electron system at high magnetic fields %%
\bibitem{Ye02} P. D. Ye, L. W. Engel, D. C. Tsui, R. M. Lewis, L. N. Pfeiffer, and K. West, Phys. Rev. Lett. {\bf 89}, 176802 (2002).

%% Evidence of two different solid phases of 2D electrons in high magnetic fields %%
\bibitem{Chen04} Y. P. Chen, R. M. Lewis, L. W. Engel, D. C. Tsui, P. D. Ye, Z. H. Wang, L. N. Pfeiffer, and K. W. West, Phys. Rev. Lett. {\bf 93}, 206805 (2004).

%% Pinning mode resonances of new phases of 2D electron systems in high magnetic fields %%
\bibitem{Sambandamurthy06} G. Sambandamurthy, Z. H. Wang, R. M. Lewis, Y. P. Chen, L. W. Engel, D. C. Tsui, L. N. Pfeiffer, and K. W. West, Solid State Commun. {\bf 140}, 100 (2006).

%% Melting of a 2D quantum electron solid in high magnetic field %%
\bibitem{Chen06} Y. P. Chen, G. Sambandamurthy, Z. H. Wang, R. M. Lewis, L. W. Engel, D. C. Tsui, P. D. Ye, L. N. Pfeiffer, and K. W. West, Nat. Phys. \textbf{2}, 452 (2006).


%% Shayegan reviews %%
\bibitem{Shayegan_Review98} For a review, see M. Shayegan, pp. 343-383 in {\it Perspectives in Quantum Hall Effects}, edited by S. Das Sarma and A. Pinczuk (Wiley, New York, 1998).

\bibitem{Shayegan_Review06} For a review, see M. Shayegan, pp. 31-60 in {\it High Magnetic Fields: Science and Technology}, Vol. 3, edited by F. Herlach and N. Miura (World Scientific, Singapore, 2006).

%% Composite Fermions waltz to the tune of a Wigner crystal %%
\bibitem{Liu14} Y. Liu, H. Deng, M. Shayegan, L. N. Pfeiffer, K. W. West, and K. W. Baldwin, arXiv:1410.3435 (2014).

%%%%%%%%%%%%%%%% Theoretical studies on the WC %%%%%%%%%%%%%%%%%%

\bibitem{Maki83} K. Maki and X. Zotos, Phys. Rev. B {\bf 28}, 4349 (1983).

\bibitem{Lam84} P. K. Lam and S. M. Girvin, Phys. Rev. B {\bf 30}, 473 (1984).

%% Hartree-Fock approximation for the "CDW" state %%
\bibitem{Levesque84} D. Levesque, J. J. Weis, and A. H. MacDonald, Phys. Rev. B {\bf 30}, 1056 (1984).

%% An improvement to Lam-Grivin results %%
\bibitem{Esfarjani90} K. Esfarjani and S. T. Chui, Phys. Rev. B {\bf 42}, 10758 (1990).

%% Collective modes of the two-dimensional WC in a strong magnetic field %%
\bibitem{Cote91} R. C\^{o}t\'{e} and A. H. MacDonald, Phys. Rev. B {\bf 44}, 8759 (1991).

%% Laughlin-Jastrow-correlated WC in a strong magnetic field %% 
\bibitem{Yi98} H. Yi and H. A. Fertig, Phys. Rev. B {\bf 58}, 4019 (1998). 

%% Hamiltonian theory of the CF WC: predicting 4CFC near nu=1/5 %%
\bibitem{Narevich01} R. Narevich, G. Murthy, and H. A. Fertig, Phys. Rev. B {\bf 64}, 245326 (2001). 

%% ED study predicting the onset filling factor for the WC to be 1/7%% 
\bibitem{Yang01} K. Yang, F. D. M. Haldane, and E. H. Rezayi, Phys. Rev. B {\bf 64}, 081301(R) (2001). 

%% DMRG study for "Ground state phase diagram of 2D electrons in high magnetic field" %%
\bibitem{Shibata03} N. Shibata and D. Yoshioka, J. Phys. Soc. Jpn. {\bf 72}, 664 (2003).

%% Study on the WC-vs-CF liquid phase diagram emphasizing that a small change in the inter-CF interaction could stabilize he CF liquid in competition to WC %%
\bibitem{Mandal03} S. S. Mandal, M. R. Peterson, and J. K. Jain, Phys. Rev. Lett. {\bf 90}, 106403 (2003). 

%% Perturbative scheme based on the correlated basis functions of CF %%
\bibitem{Jeon04} G. S. Jeon, C. C. Chang, and J. K. Jain, J. Phys. Cond. Mat. {\bf 16} L271 (2004); Phys. Rev. B {\bf 69}, 241304(R) (2004).

%% Microscopic verification of topological electron-vortex binding in the LLL crystal state showing the goodness of the CFC wave function in a quantum-dot-like configuration with L projection %%
\bibitem{Chang05} C.-C. Chang, G. S. Geon, and J. K. Jain, Phys. Rev. Lett. {\bf 94}, 016809 (2005).

%% Path integral Monte Carlo study on the phase boundary between the FQH liquid and the WC %%
\bibitem{He05} W. J. He, T. Cui, Y. M. Ma, C. B. Chen, Z. M. Liu, and G. T. Zou, Phys. Rev. B {\bf 72}, 195306 (2005).

%% Competition between CF crystal and liquid orders at 1/5: A follow-up study of Chang05 %%
\bibitem{Chang06} C.-C. Chang, C. T\"{o}ke, G. S. Jeon, and J. K. Jain, Phys. Rev. B {\bf 73}, 155323 (2006). % A follow-up study of Chang05

%% CF solid and liquid states in two component quantum dots %%
\bibitem{Shi07} C. Shi, G. S. Jeon, and J. K. Jain, Phys. Rev. B {\bf 75}, 165302 (2007).

%% Static and dynamic properties of type-II CF Wigner crystals: computation of the shear modulus and the magnetophonon and magnetoplasmon dispersions of CFC %%
\bibitem{Archer11} A. C. Archer and J. K. Jain, Phys. Rev. B {\bf 84}, 115139 (2011).

%%%%%%%%%%%%%%%% Thomson problem and CFC %%%%%%%%%%%%%%%%%%
\bibitem{Archer13} A. C. Archer, K. Park, and J. K. Jain, Phys. Rev. Lett. {\bf 111}, 146804 (2013).

\bibitem{Thomson04} J. J. Thomson, Philos. Mag. {\bf 7}, 237 (1904).
%%%%%%%%%%%%%%%%%%%%%%%%%%%%%%%%%%%%%%%%%%%%%%%%%%%%%%%%%


\bibitem{Bonsall77} L. Bonsall and A.~A. Maradudin, Phys. Rev. B {\bf 15}, 1959 (1977).


%%%%%%%%%%%%%% Measurements of the compressibility %%%%%%%%%%%%%%%%%%%%%

%% Compressibility of the two-dimensional electron gas: Measurements of the zero-field exchange energy and fractional quantum Hall gap %%
\bibitem{Eisenstein94} J. P. Eisenstein, L. N. Pfeiffer, and K. W. West, Phys. Rev. B {\bf 50}, 1760 (1994).

%% Density of states and zero Landau level probed through capacitance of graphene %%
\bibitem{Ponomarenko10} L. A. Ponomarenko, R. Yang, R. V. Gorbachev, P. Blake, A. S. Mayorov, K. S. Novoselov, M. I. Katsnelson, and A. K. Geim, Phys. Rev. Lett. {\bf 105}, 136801 (2010)

%% Interaction phenomena in graphene seen through quantum capacitance %%
\bibitem{Yu13} G. L. Yu, R. Jalil, B. Belle, A. S. Mayorov, P. Blake, F. Schedin, S. V. Morozov, L. A. Ponomarenko, F. Chiappini, S. Wiedmann, U. Zeitler, M. I. Katsnelson, A. K. Geim, K. S. Novoselov, and D. C. Elias, Proc. Natl. Acad. Sci. U.S.A. {\bf 110}, 3282 (2013). 

%% Local Compressibility Measurements of Correlated States in Suspended Bilayer Graphene %% 
\bibitem{Martin10}  J. Martin, B. E. Feldman, R. T. Weitz, M. T. Allen, and A. Yacoby, Phys. Rev. Lett. {\bf 105}, 256806 (2010). 

%% Unconventional Sequence of Fractional Quantum Hall States in Suspended Graphene %%
\bibitem{Feldman12} B. E. Feldman, B. Krauss, J. H. Smet, and A. Yacoby, Science {\bf 337}, 1196 (2012). 

%% Fractional Quantum Hall Phase Transitions and Four-Flux States in Graphene %%
\bibitem{Feldman13} B. E. Feldman, A. J. Levin, B. Krauss, D. A. Abanin, B. I. Halperin, J. H. Smet, and A. Yacoby, Phys. Rev. Lett. {\bf 111}, 076802 (2013). 

%%%%%%%%%%%%%%%%%%%%%%%%%%%%%%%%%%%%%%%%%%%%%%%%%%%%%%


\bibitem{Mahan} G. D. Mahan, \textit{Many-Particle Physics} (Plenum Press, New York, 1990).

\bibitem{Giuliani} G. F. Giuliani and G. Vignale, \textit{Quantum Theory of the Electron Liquid} (Cambridge University Press, New York, 2005).


%%%%%%%%%%%%%%%%%% Experimental indications of CFC states %%%%%%%%%%%%%%%%%%%%

%% Microwave spectroscopic observation of distinct electron solid phases in wide quantum wells %%
\bibitem{Hatke14} A. T. Hatke, Y. Liu, B. A. Magill, B. H. Moon, L. W. Engel, M. Shayegan, L. N. Pfeiffer, K. W. West, and K. W. Baldwin, Nat. Commun. {\bf 5}, 4154 (2014).

%% FQHE and WC of interacting CFs %%
\bibitem{Liu14_PRL} Y. Liu, D. Kamburov, S. Hasdemir, M. Shayegan, L. N. Pfeiffer, K. W. West, and K. W. Baldwin, Phys. Rev. Lett. {\bf 113}, 246803 (2014).

%% Theta-function handbook %%
\bibitem{NIST_Handbook} See Chapter 20 {\it Theta Functions} written by W. P. Reinhardt and P. L. Walker in F. W. J. Olver, D. W. Lozier, R. F. Boisvert, and C. W. Clark, Ed., {\it NIST Handbook of Mathematical Functions} (Cambridge University Press, New York, USA, 2010); Online companion in http://dlmf.nist.gov/20.



\end{references}
\end{document}